# Fabrication of three-dimensional microdisk resonators in calcium fluoride by femtosecond laser micromachining

Jintian Lin[1,3], Yingxin Xu[2], Jialei Tang[1,3], Nengwen Wang[1], Jiangxin Song[1,3], Fei He[1], Wei Fang[2,*], and Ya Cheng[1,**]

[1]*State Key Laboratory of High Field Laser Physics, Shanghai Institute of Optics and Fine Mechanics, Chinese Academy of Sciences, Shanghai 201800, China*
[2] *State Key Laboratory of Modern Optical Instrumentation, Department of Optical Engineering, Zhejiang University, Hangzhou 310027, China*
[3]*University of Chinese Academics of Science, Beijing 100049, China*
[*]*wfang08@zju.edu.cn*
[**]*ya.cheng@siom.ac.cn*

**Abstract:**

We report on fabrication of on-chip calcium fluoride ($CaF_2$) microdisk resonators using water-assisted femtosecond laser micromachining. Focused ion beam (FIB) milling is used to create ultra-smooth sidewalls. The quality (Q)-factors of the fabricated microresonators are measured to be $4.2 \times 10^4$ at wavelengths near 1550 nm. The Q factor is mainly limited by the scattering from the bottom surface of the disk whose roughness remains high due to the femtosecond laser micromachining process. This technique facilitates formation of on-chip microresonators on various kinds of bulk crystalline materials, which can benefit a wide range of applications such as nonlinear optics, quantum optics, and chip-level integration of photonic devices.

## 1. Introduction

Whispering-gallery-mode (WGM) microresonators have now been widely used in nonlinear optics [1-3], quantum physics [4-6], and biosensing [7, 8] owing to their unique capability of confining light efficiently in a tiny volume for long periods of time by total internal reflection [9]. Depending on the requirements of different applications, WGM microresonators have been realized in various kinds of materials including amorphous glass, crystal, polymer, liquid, etc [10-13]. Although fabrication of microresonators in amorphous glass materials has enabled realization of high-Q factors up to $\sim 10^9$ level [14], many nonlinear optical effects can only be achieved in crystalline materials as crystals can provide the desired optical properties such as high nonlinearity, wide window of transparency, very low intrinsic absorption, and so on. Today, most high-Q crystal WGM microresonators are fabricated by mechanical polishing [15]. Based on this approach, integration of small WGM microresonators with diameters of a few microns to a few tens of microns into a chip is challenging due to the small sizes of the structures and the required additional post-assembling process. As a first attempt to overcome this issue, here we demonstrate fabrication of on-chip $CaF_2$ microdisk resonators using femtosecond laser micromachining. Particularly, it should be noticed that previously, high-Q on-chip microresonators fabricated in both passive and active glass materials have been achieved using femtosecond laser micromachining [16-18]. Fabrication of amorphous glass-based high-Q microresonators can be easier because smoothing of the sidewalls can be achieved by a surface reflow process induced by either $CO_2$ laser reflow or heat reflow. However, crystalline microresonators cannot be treated similarly, due to the formation of randomly oriented crystallites on the surface [19]. In the current experiment, smoothing of sidewalls of $CaF_2$ microdisks is achieved using focused ion beam (FIB) milling [20], leading to a Q-factor of $4.2 \times 10^4$.

## 2. Fabrication of microdisk resonators by femtosecond laser micromachining

Commercially available Z-cut $CaF_2$ substrates with a thickness of 1mm were used as the material platforms for fabrication of the microdisks. Both the upper and bottom surfaces of the substrate were polished. As illustrated in Fig. 1, the procedures for fabrication of $CaF_2$ microdisk resonators are: (1) femtosecond laser selective ablation of a $CaF_2$ substrate immersed in water from its backside to form a freestanding disk-shaped microstructure; and (2) smoothing of the sidewall of the disk-shaped microstructure by FIB milling to create the ultra-smooth sidewall of the microdisk.

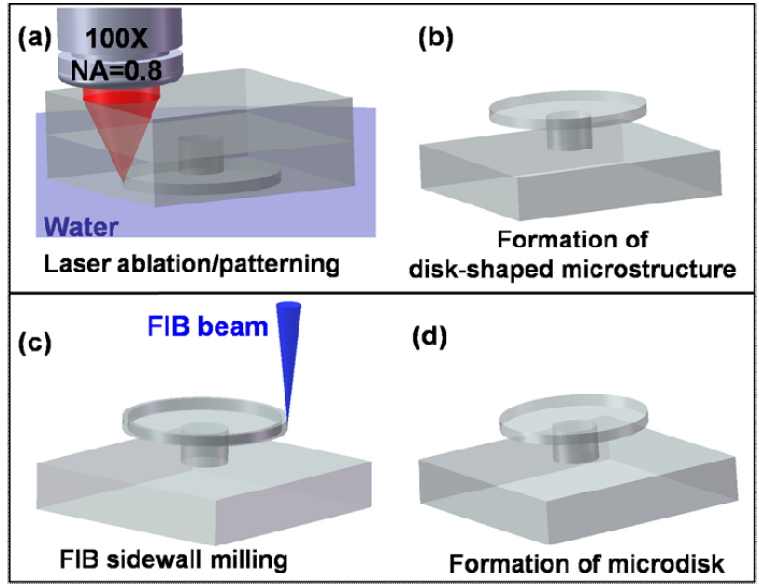

**Fig. 1** Procedures of fabrication of $CaF_2$ microdisk by femtosecond laser ablation assisted by water and followed by FIB milling.

The home-made femtosecond laser micromachining workstation consisted of a Ti: sapphire femtosecond laser source (Coherent, Inc., center wavelength: ~800nm, pulse width: ~40fs, repetition rate: 1 kHz) coupled to a microscope system (objective lens with a numerical aperture of 0.8). The average power of the femtosecond laser was adjusted to be 1 mW using a variable density filter. The $CaF_2$ substrate was fixed on the computer-controlled XYZ translation stage with a resolution of 1μm. In our experiment, the ablation front always contacted with deionized water to remove the ablation debris. The material was selectively removed to form the freestanding disk-shaped microstructures by continuously layer-by-layer annular scanning the focal spot (~1 μm) in the region surrounding the predesigned micro-disk structure. 1.4-μm interval between the adjacent layers was adopted to ensure sufficient overlap between the adjacent ablation layers. After the laser ablation, the disk-shaped microstructure with a diameter of 88 μm, a thickness of ~15 μm, which was supported by a thin pillar, was formed. The entire ablation process took ~120 min. Figure 2(a) shows the scanning electron microscope (SEM) image of the disk-shaped microstructure. It can be seen that the roughness of the sidewall was high as a result of laser ablation. Although the bottom surface cannot be directly examined by SEM, we expect that it should have a surface roughness similar to that of the sidewalls, because the same ablation conditions were used for creating both of them. The roughness of the sidewalls can cause significant scattering loss, thus it must be reduced for promoting the Q-factor.

## 3. Smoothing of the sidewall of the fabricated microdisk

Although $CO_2$ laser induced surface reflow has frequently been used to create ultra-smooth silica microresonators, this method is usually not applicable for most crystalline materials. For example, in the current experiment, $CaF_2$ is transparent to $CO_2$ laser beam; for other crystalline materials that could absorb $CO_2$ laser, the surface thermal reflow would create randomly oriented crystallites due to rapid heating and cooling which act as scattering centers.

Previously, FIB has been employed in precise fabrication of millimeter-scale optical resonators to shape the perimeters of microdisks by milling the edges [20-22]. Unlike these investigations, in our experiment, FIB was used to mill the edge of the $CaF_2$ microdisk to attain ultra-smooth sidewalls. The fabricated sample was mounted on a short brass pedestal after sputter coated with a ~20 nm layer of gold to reduce charge collection during the FIB milling. Then the sample was installed in a dual-beam SEM/FIB system (Carl Zeiss, Model: AURIGA) with the top surface of the microdisk facing the ion beam. The FIB milling was operated twice, beginning with a coarse milling and followed by a fine one for further improving the surface smoothness. In the coarse milling, a 30 keV ion beam with a beam current of 30 nA was used to polish the sidewall. After the initial coarse milling, the microdisk showed a much smoother sidewall, as illustrated in Fig. 2(b). However, magnified SEM image (Fig. 2(d)) indicates the existence of fine vertical ripples on the sidewall oriented parallel to the direction of the ion beam. These fine ripples can be removed by performing an additional FIB milling with finer beam sizes and lower beam currents (16nA). Figure 2(c) shows an SEM image of the microdisk undergone the fine milling, whose diameter was slightly reduced to ~79 μm. The smooth sidewall is further evidenced by its close-up view in Fig. 2(e). The result also indicates that femtosecond laser micromachining followed by FIB milling can enable fabrication of the high quality microdisks in crystalline materials, not only limited to $CaF_2$.

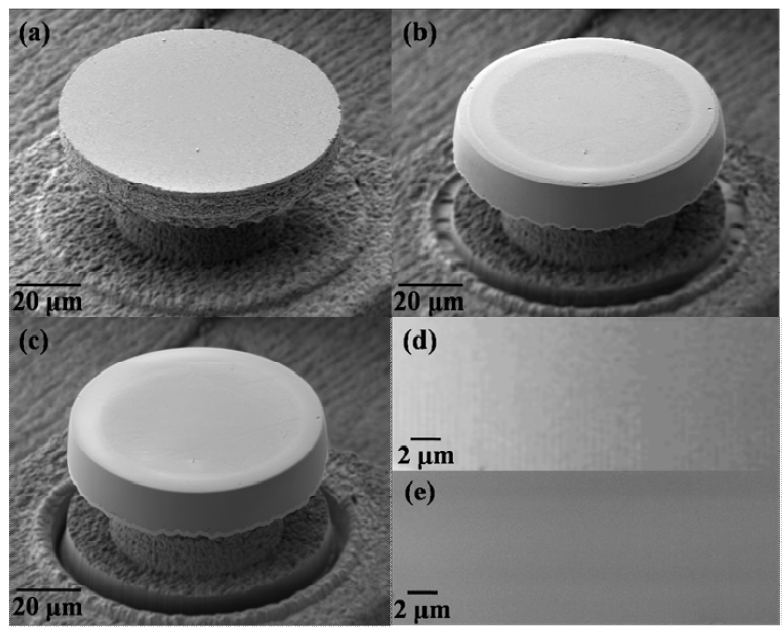

**Fig. 2** SEM images of a $CaF_2$ microdisk fabricated by femtosecond laser micromachining **(a)** before, **(b)** after initial coarse FIB milling, and **(c)** after fine milling. **(d)** and **(e)** show the enlarged SEM images of sidewalls of the microdisk before and after the fine milling, respectively. The fine ripples in **(d)** disappear with the fine milling, as evidenced in **(e)**.

## 4. Characterization of microresonator

Because the refractive index of the $CaF_2$ is close to that of a silica optical fiber, the fiber taper coupling method [23] was chosen to measure the Q-factor and characterize the transmission spectrum of the CaF$_2$ microdisk, as shown in Fig. 3(a). The fiber-microdisk coupling system consisted of a tunable laser source (Agilent, Inc., Model: 81600B), a transient optical power meter (Agilent, Inc., Model: N7744A), a polarization synthesizer (Agilent, Inc., Model: N7786B), and a piezo stage (PI, Inc., Model: E-463). The Alight sweeping from 1546 to 1560 nm from a tunable laser source was sent into the fiber taper coupled with the microdisk. The fiber taper had a waist diameter of approximately ~1 μm, enabling excitation of whispering gallery modes in the microdisk resonator by evanescent wave coupling. The transmission spectrum was measured using the transient optical power meter. Dual CCD cameras were used to monitor, from both the horizontal and the vertical directions, the relative position between the fiber taper and the microdisk. The microdisk was fixed on the piezo stage with a positioning resolution of 50 nm in the XYZ directions. The critical coupling was realized by carefully adjusting the relative position between the fiber tape and the microdisk. The resonance transmission spectrum of the fiber taper coupled to the microdisk is presented in Fig. 3(c). The azimuthal free spectral range (FSR) was measured to be 6.43 nm. The azimuthal FSR can also be estimated by the relation $\Delta\lambda=\lambda^2/(n\times C)$, where $\lambda$ is the emission wavelength, $n$ is the effective index of microdisk, and $C$ is the circumference of the microdisk. Such an approximation is commonly used for microdisk resonators whose modes are located near the circumference. For $n$=1.43 and $C$=276.46μm, the estimated value of the azimuthal FSR is approximately 6.10 nm when $\lambda$ is 1552.94 nm, which is close to the measured azimuthal FSR. Fig. 3(d) shows an individual WGM located at 1546.51 nm wavelength with a Lorentzian-shaped dip. The linewidth obtained by a Lorentzian fitting is 36.74 pm, as shown by the red curve Fig. 3(d). The Q factor for the microdisk is then calculated to be $4.2\times10^4$.

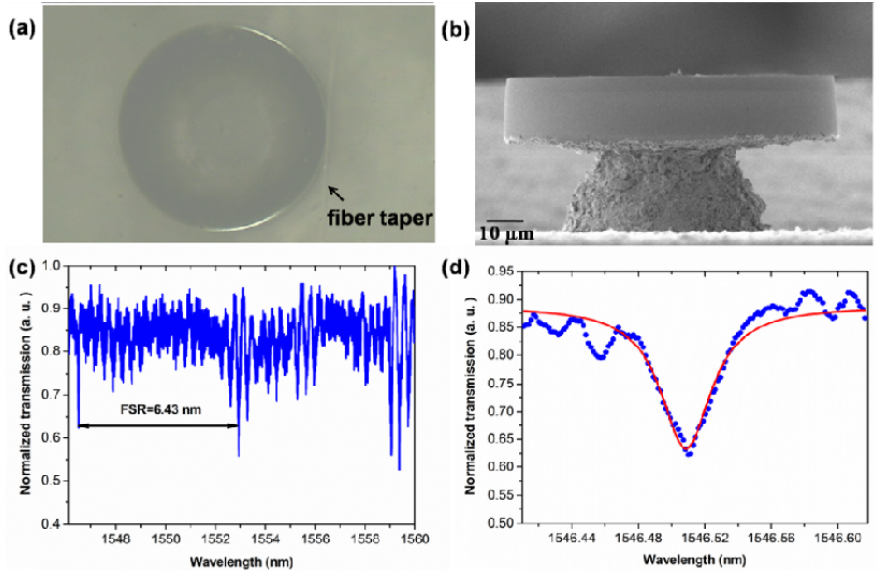

**Fig. 3 (a)** Optical micrograph of the microdisk coupled with a fiber taper (top view). **(b)** SEM image of the fabricated microdisk (side view), showing the tilting of the sidewall. **(c)** Transmission spectrum of the fiber taper coupled with the microdisk. **(d)** Lorentzian fit (red solid line) of measured spectrum around the resonant wavelength at 1546.51nm (blue dotted line), showing a Q factor of $4.2\times10^4$.

## 5. Discussion and conclusions

As shown in Fig. 2, a smooth surface has been achieved by performing the FIB milling for the sidewalls of the microdisk fabricated by femtosecond laser micromachining. However, even with such a smooth sidewall, the Q-factor measured in our experiment is still less than $10^5$. It is critical to clarify the major cause behind the low Q-factor for further improvement in the future. As one may have noticed in Fig. 3(b), the sidewall of the microdisk was slightly tilted with a tilting angle of 3.8° with respect to the normal direction of the upper surface, which was naturally formed during the FIB milling process. Although it has been shown in previous experiments that a wedge-shaped edge of the microdisk [24] can be helpful for achieving ultrahigh Q-factors due to the isolation of modes from the perimeter, in such cases, the bottom surfaces must be very smooth for avoiding scattering loss of the light propagating along the tip of the wedge. Unfortunately, in our case, the bottom surface formed by femtosecond laser micromachining was quite rough, thereby limiting the achievable Q-factor. This problem can possibly be solved by improving the FIB milling technique to reduce or even eliminate the conical effect or by improving the bottom surface quality of the microdisks with coating or chemical polishing.

To summarize, on-chip $CaF_2$ microdisk resonators have been fabricated by femtosecond laser micromachining, followed by an FIB milling for smoothing the sidewalls. The measured Q-factor of the microresonators reaches $4.2 \times 10^4$. It should be noted that with the smaller diameters of the microresonators, larger FSR values can be achieved. Further improvement of Q factor is possible by either reducing the titling angle of the sidewall or improving the smoothness of the bottom surface. The technique may eventually enable chip-level integration of crystal-based microresonators.

**Acknowledgements**

We thank Prof. Min Qiu from Zhejiang University for their kind help. The work is supported by National Basic Research Program of China (Nos. 2014CB921300), NSFC (Nos. 61275205, 11104245, 61108015, 61008011, 11174305, 11104294, and 61205209), the Program of Shanghai Subject Chief Scientist (11XD1405500), and the Fundamental Research Funds for the Central Universities.

**References**

1. S. M. Spillane, T. J. Kippenberg, and K. J. Vahala, "Ultralow-threshold Raman laser using a spherical dielectric microcavity," Nature **415**, 621 (2002).
2. P. Del'Haye, A. Schliesser, O. Arcizet, T. Wilken. R. Holzwarth, and T. J. Kippenberg, "Optical frequency comb generation from a monolithic microresonator," Nature **450**, 1214 (2007).
3. T. Carmon and K. J. Vahala, "Visible continuous emission from a silica microphotonic device by third-harmonic generation," Nat. Phys. **3**, 430 (2007).
4. D. W. Vernooy, A. Furusawa, N. Ph. Georgiades, V. S. Ilchenko, and H. J. Kimble, "Cavity QED with high-Q whispering gallery modes," Phys. Rev. A **57**, R2293 (1998).

5. T. Aoki, B. Dayan, E. Wilcut, W. P. Bowen, A. S. Parkins, T. J. Kippenberg, K. J. Vahala, and H. J. Kimble, "Observation of strong coupling between one atom and a monolithic microresonator," Nature **443**, 671 (2006).

6. E. Peter, P. Senellart, D. Martrou, A. Lemaître. J. Hours, J. M. Gérard, and J. Bloch, "Exciton-photon strong-coupling regime for a single quantum dot embedded in a microcavity," Phys. Rev. Lett. **95**, 067401 (2005).

7. F. Vollmer and S. Arnold, "Whispering-gallery-mode biosensing: label-free detection down to single molecules," Nat. Methods **5**, 591 (2008).

8. J. Zhu, S. K. Ozdemir, Y.-F. Xiao, L. Li, L. He, D.-R. Chen, and L. Yang, "On-chip single nanoparticle detection and sizing by mode splitting in an ultrahigh-Q microresonator," Nat. Photonics **4**, 46 (2010).

9. S. L. McCall, A. F. Levi, R. E. Slusher, S. J. Pearton, and R. A. Logan, "Whispering-gallery mode microdisk lasers," Appl. Phys. Lett. **60**, 289 (1992).

10. D. K. Armani, T. J. Kippenberg, S. M. Spillane, and K. J. Vahala, "Ultra-high-Q toroid microcavity on a chip," Nature **421**, 925 (2003).

11. I. S. Grudinin, A. B. Matsko, A. A. Savchenkov, D. Strekalov, V. S. Ilchenko, L. Maleki, "Ultra high Q crystalline microcavities," Opt. Commun. **265**, 33 (2006).

12. T. Grossmann, S. Schleede, M. Hauser, T. Beck, M. Thiel, G. v. Freymann, T. Mappes, and H. Kalt, "Direct laser writing for active high-Q polymer microdisk on silicon," Opt. Express **19**, 11451 (2011).

13. H.-M. Tzeng, k. F. Wall, M. B. Long, and R. K. Chang, "Laser emission from individual droplets at wavelengths corresponding to morphology-dependent resonances," Opt. Lett. **9**, 499 (1984).

14. M. L. Gorodetsky, A. A. savchenkov, and V. S. Ilchenkov, "Ultimate Q of optical microsphere resonators," Opt. Lett. **21**, 453 (1996).

15. A. A. Savchenkov, V. S. Ilchenko, A. B. Matsko, and L. Maleki, "Kilohertz optical resonances in dielectric crystal cavities," Phys. Rev. A **70**, 051804(R) (2004).

16. J. Lin, Y. Xu, J. Song, B. Zeng, F. He, H. Xu, K. Sugioka, W. Fang, and Y. Cheng, "Low-threshold whispering-gallery-mode microlasers fabricated in a Nd:glass substrate by three-dimensional femtosecond laser micromachining," Opt. Lett. **38**, 1458 (2013).

17. J. Lin, S. Yu, Y. Ma, W. Fang, F. He, L. Qiao, L. Tong, Y. Cheng, and Z. Xu, "On-chip three-dimensional high-Q microcavities fabricated by femtosecond laser direct writing," Opt. Express **20**, 10212 (2012).

18. K. Tada, G. A. Cohoon, K. Kieu, M. Mansuripur, and R. A. Norwood, "Fabrication of high-Q microresonators using femtosecond laser micromachining," IEEE Photo. Tech. Lett. **25**, 430 (2013).

19. V. S. Ilchenko, A. A. Savchenkov, A. B. Matsko, and L. Maleki, "Nonlinear optics and crystalline whispering gallery mode cavities," Phys. Rev. Lett. **92**, 043903 (2004).

20. C. F. Wang, Y-S. Choi, J. C. Lee, E. L. Hu, J. Yang, and J. E. Butler, "Observation of whispering gallery modes in nanocrystalline diamond microdisks," Appl. Phys. Lett. **90**, 081110 (2007).

21. L. A. M. Barea, F. Vallini, A. R. Vaz, J. R. Mialichi, and N. C. Frateschi, "Low-roughness active microdisk resonators fabricated by focused ion beam," J. Vac. Sci. Technol. **B 27**, 2979 (2009).

22. J. R. Mialichi, L. A. M. Barea, P. L. d. Souza, R. M. S. Kawabata, M. P. Pires, and N. C. Frateschi,, "Resonance modes in InAs/InGaAlAs/InP quantum dot microdisk resonators," ECS Trans. **31**, 289 (2010).

23. A. Serpengüzel, S. Arnold, and G. Griffel, "Excitation of resonances ofmicrospheres on an optical fiber," Opt. Lett. **20**, 654 (1995).

24. T. J. Kippenberg, S. M. Spillane, D. K. Armani, and K. J. Vahala, "Fabrication and coupling to planar high-Q silica disk microcavities," Appl. Phys. Lett. **83**, 797 (2003).